# The Algorithmic Unconscious:
# Structural Mechanisms and Implicit Biases in Large Language Models

Philippe Boisnard
Paris 8 - Paragraphe-CITU

*Abstract* :

This article introduces the concept of the *algorithmic unconscious* to designate the set of structural determinations that operate within large language models (LLMs) without being accessible either to the model's own reflexivity or to that of its users. In contrast to approaches that reduce AI bias solely to dataset composition or to the projection of human intentionality, we argue that certain biases emerge directly from the technical mechanisms themselves-tokenization, attention, statistical optimization, and alignment procedures.

Based on a comparative analysis of tokenization across a corpus of parallel sentences, we show that Arabic languages (Modern Standard Arabic and Maghrebi dialects) undergo a systematic inflation in token count relative to English, with ratios ranging from 1.6x to nearly 4x depending on the infrastructure (OpenAI, Anthropic, SentencePiece/Mistral). This over-segmentation constitutes a measurable infrastructural bias that mechanically increases inference costs, constrains access to the contextual space, and affects the attentional weighting of representations.

We relate these empirical findings to three additional structural mechanisms-causal bias (correlation ≠ causation), the erasure of minoritized features (dimensional collapse), and normative biases induced by safety alignment-in order to show how LLMs reproduce and amplify linguistic and cultural hierarchies.

Finally, we propose a framework for a technical clinic of models, grounded in the audit of tokenization, latent space topology, and alignment systems, as a necessary condition for a critical appropriation of AI infrastructures.

It has been widely acknowledged for more than a decade that machine learning and deep learning systems produce biases as a function of the composition of their training datasets. As early as 2016, it was noted that *"blind application of machine learning can amplify biases present in the data. This danger is particularly acute with word embeddings, a popular framework for vector-based representations of textual data used across many machine learning and natural language processing tasks"* (Bolukbasi et al., 2016).

This form of algorithmic causality—embedded both in learning procedures and in the construction of latent spaces—is what we propose to interrogate through the concept of the algorithmic unconscious.

What we call the algorithmic unconscious is not a psychological metaphor projected onto machines. Rather, it designates everything that operates within AI systems without being thought: what is embedded in technical layers, data selection choices, cleaning procedures, tokenization schemes, filtering pipelines, and alignment mechanisms. It is "unconscious" in a double sense. First, these processes are not accessible to users' reflexive awareness. Second, the AI system itself lacks the capacity—when producing outputs—to interrogate or reflect upon the biases constitutive of its own operation.

While this notion partially overlaps with Luca Possati's use of the term *algorithmic unconscious*, our approach diverges in a crucial respect. Possati primarily analyzes the projection of the human unconscious onto machines, whereas we argue for the existence of an unconscious intrinsic to the model itself—that is, structural biases that emerge directly from technical mechanisms such as tokenization, attention, and alignment. This requires us to examine algorithmic causalities as such, rather than reducing bias to human intention alone.

Our analysis therefore investigates how internal learning processes can generate bias, including:

— contextual weighting of tokens through attention layers;

— causal bias in learning, insofar as neural networks do not learn causes but associative inductive contexts;

— the erasure of minoritized features through statistical preference (*dimensional collapse*);

— and the formation of hidden biases induced by safety alignment procedures in contemporary LLMs.

We further show that these biases are not uniform across LLMs but may be intrinsic to specific encoding logics. A comparison between Mistral and ChatGPT illustrates this point: although their tokenization strategies appear similar at first glance (SentencePiece vs. BPE), Mistral produces significantly finer segmentation, particularly for non-English llanguages that are not well covered by English-dominant tokenization vocabularies.

Making the algorithmic unconscious visible is necessary because it has become structural to knowledge production as LLMs such as ChatGPT are increasingly used in research, education, and everyday writing practices (Kannan, 2024). While much attention has been devoted to content-level bias, it is equally crucial to foreground the algorithmic mechanisms underlying these usages. Through concrete examples drawn from Arabic and Darija, we demonstrate multiple levels of algorithmic bias and show that far from being neutral, these mechanisms have social and economic consequences, especially for speakers of non-majority languages. At a time when AI systems are being rapidly and often uncritically adopted—particularly by students—it is essential to expose the foundations and effects of these algorithmic biases for cultures and languages that were not central to the parameterization of contemporary generative AI systems.

## 1. Contextual Weighting and Representational Inequalities Induced by Tokenization

*(Attention Mechanisms and Internal Hierarchies)*

To properly understand what occurs when interacting with an LLM via prompts, it is necessary to schematically describe the stages through which a prompt acquires algorithmic meaning. What the model processes is not semantic meaning in an intellectual sense, but an algorithmic one—produced by the vector transformation of an utterance and its insertion into a vector field (context).

The process begins with the prompt.

The prompt is segmented into tokens (numerical identifiers) by a tokenizer. These tokens are then embedded, i.e., mapped into a vector space. From there, the model applies attention mechanisms, most notably self-attention.

The algorithmic unconscious begins precisely in the interaction between tokenization and attention layers. Models assign markedly different weights to tokens depending on their frequency, morphological structure, granularity (subwords), and occurrence within the training corpus. This process generates an internal geography of tokens, in which certain languages, morphologies, or dialects—most notably English—occupy central or referential positions, while others remain peripheral.

When a prompt is written, once it has been tokenized and embedded, it is no longer interpreted according to the "natural" context of its enunciation. Instead, it is translated into and constrained by a vector field that becomes its context of analysis. For mainstream generative AI applications—starting with ChatGPT—this vector field is predominantly shaped by the English language and by a largely Western, and more specifically North American, cultural background.

As a result, disparities and unconscious distortions of meaning emerge as early as the tokenization stage. SentencePiece-based tokenizers (as used in Mistral) tend to produce longer units for high-resource languages—resulting in fewer tokens—while generating more fragmented tokenizations for low-resource or under-resourced languages. This leads to more unstable segmentations for rare or marginalized languages. Consequently, the tokenization context derived from model design and parameterization enables more precise and efficient interaction with majority languages than with minority ones, a pattern that extends to cultural content as well.
As Teklehaymanot and Nejdl (2025) observe:

> *« While multilingual LLMs (mLLMs) theoretically enable cross-lingual knowledge transfer from high-resource to low-resource languages through shared representations, substantial performance disparities persist across linguistically diverse populations. Recent analyses of code-switching datasets reveal systemic biases toward English and a lack of sociolinguistic representativeness in data collection and annotation, reflecting broader inequities in multilingual modeling. »*

When considering Arabic, it becomes evident that models such as SentencePiece/Mistral often segment words very finely during tokenization—sometimes down to the character level (see Appendices 1–2)—in line with findings reported by Bari et al. (2024). This fine-grained segmentation requires, on average, increased sampling to stabilize a word's possible meaning (Alyafeai et al., 2024). The effect is even more pronounced for Darija, which is doubly minoritized: it is a dialect—with multiple variants—of a language that is itself underrepresented. Researchers working on Atlas-Chat note that LLMs "often neglect underrepresented languages" by granting them access only through significantly higher token fertility compared to majority languages (Shang et al., 2024).

Our comparative analysis of tokenization across a corpus of parallel sentences reveals a systematic inflation in token counts for Arabic languages, with ratios ranging from 1.6× (OpenAI) to nearly 4× (SentencePiece/Mistral) relative to English. This inflation is not uniform across infrastructures and exposes an implicit hierarchy among languages at the level of subword segmentation (see Appendix 2).

Such over-segmentation (fertility) in minority languages like Arabic has direct economic implications for LLM usage. The training cost of a model can be multiplied by up to four for a language with doubled fertility (Petrov et al., 2023). This cost is subsequently transferred to users, who consume more context tokens for equivalent queries in languages with less efficient tokenization.

Tokenization disparities also affect the distribution of attention within the model. Attention in AI systems, like human attention, is not neutral. It structurally materializes a statistical unconscious shaped by both the corpus and the tokenization process. Some tokens become over-significant—highly recurrent and linked to multiple contexts—thereby polarizing the attentional space. Others become invisible, silenced, or reduced to noise (Alrefaie et al., 2024). This attentional disparity constitutes the first level of the algorithmic unconscious, forming a foundational component of the specific associative logic through which AI systems operate.

2. Causal Bias in Learning *(Correlations ≠ Causes)*

The second structural bias is more intuitively grasped. It rests on the observation articulated by Judea Pearl (2018) that causality does not exist from the standpoint of purely statistical induction in AI systems.

Large language models do not learn causes, but conditional distributions: *"If a given token appears in a given context, then the most probable continuation is X."* Causality cannot emerge from statistical observation alone; it requires intervention—the ability to distinguish between *seeing* and *doing*. By construction, however, LLMs observe only textual co-occurrences. While they may reproduce causal relations that are present in the data, they cannot infer causal structures. They learn that certain tokens tend to follow others, without ever accessing the causal mechanisms that generate these regularities (e.g., natural laws versus lexical field co-occurrences). This limitation is not a correctable flaw but a constitutive property of autoregressive architectures. For AI models, there is no ontologically unified reality to be experimented upon from which general laws could be inferred; their "reality" consists of data and of the accumulation of data that forms the conditional context for induction.

This produces a systemic causal bias: models tend to treat correlations as implicit causal relations, which can in turn amplify stereotypes or negative cultural associations. Such implicit causalities must be examined because they drive extrapolation processes both within majority languages and, even more critically, within minority languages. When minority languages lack sufficient contextual diversity, they are often interpreted through the causal frameworks of majority languages.

Concretely, a query formulated in Arabic, due to insufficient latent contextualization, may be interpreted through culturally heterogeneous or externally imposed causal structures—often Western in origin. These associations materialize geometrically within the latent space, producing unpredictable curvature patterns. Naous and Xu (2025) show, for example, that embeddings of concepts related to Islam in LLMs exhibit negative vector proximities (e.g., to violence or terrorism) that do not appear for analogous concepts in other cultural contexts. These associations are not intrinsically "true"; they result from representational structure, encoding choices, and neighborhood relations induced during inference.

Similarly, some models tend to interpret certain tokens as causal noise and consequently filter them—a phenomenon known as *contextual overfitting* in multilingual models. This effect stems from excessive tokenization fertility. Certain dialects are treated as noisy because available contexts are insufficient to stabilize associations among overly fragmented elements. Morphological richness, which is a linguistic asset, thus becomes a liability in computational processing and is filtered out.

The model does not understand narrative patterns or cultural causality; it induces regularities. This mechanism often leads to the involuntary essentialization of underrepresented cultures and to reductions of causal fields. Such essentialization differs from classical representational bias in datasets, as it operates more deeply and less visibly than simple overrepresentation effects (Bowker & Star, 1999; Noble, 2018). Bias is too often understood solely in terms of overrepresentation, whereas here we observe that underrepresentation itself produces essentialization. In the absence of sufficiently diverse contexts, a culture is reduced to a small set of stereotypical associations—those most frequently occurring in a dataset already shaped by majority-language contexts. Cultural complexity is flattened into a narrow schema of statistical correlations. In this sense, the algorithmic unconscious produces a form of causal reductionism.

### 3. Erasure of Minoritized Features *(Dimensional Collapse)*

The third mechanism of the algorithmic unconscious operates through a dynamic of erasure. By design, LLMs tend to privilege majority statistical regularities at the expense of minoritized forms. This phenomenon, which can be described as dimensional collapse, functions as an attractive force toward dominant modes within the latent space. Minority languages undergo dimensional collapse in LLM latent representations: their embeddings occupy a reduced, anisotropic subspace, limiting their semantic expressivity. As documented by Ethayarajh (2019) and Jing et al. (2022), this effect is amplified for underrepresented linguistic varieties such as Darija, whose embeddings tend to converge toward those of Modern Standard Arabic rather than preserving dialect-specific features.

To understand this phenomenon, it is necessary to return to the optimization logic governing model training. An LLM seeks to minimize its loss function—that is, to maximize its ability to predict the next token. From a statistical standpoint, it is more "efficient" for the model to accurately predict frequent tokens than rare ones. A marginal gain in accuracy for a token that appears millions of times carries far greater weight than an equivalent gain for a rare token.

As shown by recent work on tokenization disparities as infrastructural bias (Petrov et al., 2023; Teklehaymanot & Nejdl, 2025), models structurally favor frequent tokens according to a logic of minimal computational effort. This preference is not deliberate; it emerges from the mechanics of optimization itself. For overall performance, the model learns to sacrifice precision at the margins in favor of robustness at the center—namely, regions of high probability characterized by frequent tokens, well-represented contexts, and stable attractors in latent space. This process can be described as a topological dynamic that privileges predictive robustness over representational diversity.

What makes this mechanism particularly insidious is that it produces erasure without any intention to erase. No one programs the model to ignore minority dialects or rare cultural practices. The model, as a system, has no intentions of its own. Erasure is an emergent effect of architecture and training.

Concretely, this means that minority linguistic, stylistic, or cultural forms gradually become mathematically illegible. Their representations in latent space deteriorate, collapse into noise, or are absorbed by more stable attractors—namely, the majority forms to which they appear similar from the model's perspective.

Evaluations of Arabic LLMs clearly reveal this mechanism. As Alrefaie et al. (2024) note, North African dialects—Moroccan Darija, Algerian Darija, Tunisian Arabic—are systematically treated as noisy variants of Modern Standard Arabic. Lacking sufficient contextual diversity to stabilize these dialects as autonomous linguistic systems, the model re-normalizes them toward the form it knows best.

This forced normalization has profound consequences. It does not merely correct spelling or grammar; it erases identity markers, cultural nuances, and specific modes of expression. A Darija speaker interacting with an LLM sees their language effectively "translated" into Modern Standard Arabic—a language that is not their mother tongue, does not carry the same connotations, and does not express the same lived realities.

This phenomenon can be described by analogy with the well-known mode collapse observed in generative models such as GANs: the model ceases to produce diversity and instead collapses onto a small number of statistically "safe" modes. In the case of LLMs, this collapse operates at the conceptual and cultural level. Minoritized forms—whether languages, dialects, writing styles, or cultural references—are progressively absorbed into dominant attractors of the latent space. A query in Darija is interpreted through the prism of Modern Standard Arabic; a Maghrebi cultural reference is read through Western or Middle Eastern schemas that are more frequent in the training corpus.

This erasure mechanism can be interpreted as a form of epistemic violence in the sense articulated by Gayatri Spivak (1988): not physical violence, but violence exerted on the very conditions of representation and expression. Minority cultures are not explicitly censored; they are rendered unthinkable by the infrastructure of the system itself.

The algorithmic unconscious thus produces what may be termed computational colonialism: not an intentional domination, but a structural reproduction of linguistic and cultural hierarchies embedded within model architectures. This notion extends the analysis of data colonialism proposed by Couldry and Mejias (2022), which focuses on asymmetric data extraction and exploitation, while shifting attention toward internal computational mechanisms. It also refines technical critiques of algorithmic colonialism (Mohamed et al., 2020), which identify the reproduction of Western norms in algorithmic systems without fully explicating the underlying architectural processes.

Languages and cultures that have lacked the resources to constitute large-scale digital corpora are thus marginalized a second time—not through historical colonialism, but through algorithms that inherit and reenact its structural traces.

Finally, this erasure mechanism is further reinforced by a last level of the algorithmic unconscious: safety alignment procedures, which, under the guise of securing model usage, introduce additional biases. These biases are intentionally programmed, yet their differentiated effects across languages and cultures remain largely unexamined.

4. Formation of Hidden Biases through Safety Alignment

The fourth mechanism of the algorithmic unconscious is perhaps the most paradoxical. It arises from efforts to make models safer and aligned with human values for the benefit of users. Contemporary safety alignment—the set of techniques designed to prevent LLMs from producing harmful content—introduces a new normative unconscious, all the more insidious because it presents itself under the guise of ethical neutrality.

Safety alignment operates at least at three levels of LLM production:

a. Training data filtering.

Before learning even begins, certain content is excluded from training corpora—material deemed toxic, offensive, or simply "inappropriate" according to criteria defined by developers.

b. Reinforcement learning from preferences (RLHF/RLAIF).

After pre-training, the model is refined via reinforcement learning, where human evaluators (or AI models) label outputs as "good" or "bad," "helpful" or "harmful." The model thus learns not only what it may say, but also what it must not say.

c. Content rules and guardrails.

Real-time moderation systems filter inputs and outputs, refusing certain prompts or censoring specific responses.

As shown by Noels et al. (2025) in their study of censorship in LLMs, the desire to monitor and restrict model outputs has generated extensive debate. Critics emphasize that censorship practices raise significant ethical, social, and practical questions: Who are the actors responsible for surveillance and control? What legitimizes their authority, and who oversees their decisions?

Following Possati's analyses (2020), one might argue that safety alignment produces a genuine algorithmic superego—an internal normative instance that imposes behavioral limits on the model. Unlike the Freudian superego, which emerges through the internalization of diverse and often contradictory social norms, this algorithmic superego is programmed according to specific criteria defined by technology companies operating within particular economic, political, and cultural contexts.

Popular LLMs appear to reflect specific ideological or cultural biases: a model may readily articulate certain perspectives while responding cautiously or evasively to others. This asymmetry is not neutral; it shapes what can be thought, said, and explored through these tools. Users—especially students—who are unaware of these mechanisms may unknowingly reproduce the unconscious partitions generated by the algorithmic unconscious.

The fundamental problem is that the norms defining what is considered "safe" or "harmful" are largely derived from Western, and more specifically North American, contexts. The datasets used to train moderation systems, the human evaluators supplying preferences for RLHF, and platform content rules are all imbued with culturally specific assumptions about acceptable speech.

Shahid et al. (2025) demonstrate that colonial biases in automated moderation pipelines stem from the use of biased machine translation systems, Western-centric toxicity models, and deficient language detection tools that fail to account for cultural nuances of online harm and the evolution of languages in the Global South.

The case of Arabic and its dialects illustrates this phenomenon particularly clearly. Prompts in Arabic are subject to lower safety robustness and reduced precision compared to other languages, likely due to limited training resources and data availability (Ashraf et al., 2025).

This produces a problematic double bind. On the one hand, Arabic dialect content is underrepresented in training data, degrading model performance. On the other hand, moderation

systems—trained on biased data—tend to over-filter such content, perceiving it as potentially offensive or toxic.

Assoudi's comparative analysis of toxicity detection systems for Moroccan Darija highlights that while LLMs exhibit some sensitivity to multilingual and dialectal variation, their weakest point remains alignment with human judgments—exposing a critical limitation of relying solely on *LLM-as-a-judge* frameworks for dialectal evaluation (Assoudi, 2025). Standardized LLM pipelines thus struggle to handle Arabic dialects due to linguistic variability, the lack of standardized resources, and cultural nuances that general-purpose models often fail to capture.

The problem is systemic and rooted in the very practices used to construct moderation datasets. As Shahid et al. (2025) note, moderation systems deployed by technology companies prioritize English-speaking users in Western contexts, leaving harmful content in Global South languages largely unchecked while frequently censoring benign content in non-English languages and silencing marginalized voices. These failures are often linked to outdated resources; one industry practitioner reported relying on an old Arabic dictionary due to the scarcity of Maghrebi Arabic datasets. More strikingly, during the Arab Spring, some major social media platforms reportedly employed only two Arabic-speaking moderators—an implausibly small number given the diversity of contexts involved.

This lack of human and data resources leads to systematic moderation errors. Dialectal content—Darija, Algerian Darija, Tunisian Arabic—is frequently classified as toxic or offensive not because of its actual content, but because moderation systems lack the contextual knowledge required for accurate interpretation (Bensalem et al., 2024). Idiomatic expressions become insults; informal registers are treated as vulgarity; cultural references are flagged as suspicious content.

For Moroccan Darija in particular, challenges are amplified by the diglossic nature of the language, in which Modern Standard Arabic used in formal contexts differs significantly from spoken dialects. As a result, models trained primarily on standard Arabic perform poorly when applied to regional dialects.

General-purpose moderation systems fail especially on culturally embedded toxic content, including implicit insults, sarcasm, and culturally specific forms of aggression that are often invisible to Western-centric detection frameworks. What constitutes an insult in one cultural context may be an affectionate term in another; what appears aggressive under Western norms may be a normal conversational register elsewhere.

Safety alignment is therefore not merely protective. It introduces a latent moral bias, often Western-centric, that shapes the observable "personality" of the model. This bias operates all the more effectively because it remains invisible: presented as universal safety, it naturalizes particular cultural norms.

Minority languages and cultures are thus doubly penalized—first through underrepresentation in training data, which degrades performance, and second through over-filtering in moderation systems, which further marginalizes them by associating them with "problematic" content. Here, the algorithmic unconscious produces what can be described as structural censorship: not an explicit prohibition, but a systematic degradation of the conditions of expression for certain languages and cultures.

## 5. Toward a Technical Psychoanalysis of Models

The four mechanisms analyzed above—contextual weighting and attentional hierarchization, causal bias, erasure of minoritized features, and the formation of hidden biases through safety alignment—do not constitute isolated flaws that could simply be "fixed." Together, they form a system, an economy of the algorithmic unconscious, whose overall structure must be understood in order to design meaningful interventions.

By analyzing the algorithmic unconscious as detached from human intention—or at least operating at a distance from it—our approach diverges from that of Possati (2020, 2021). For Possati, the algorithmic unconscious primarily designates the projection of the human unconscious onto machines: through processes of projective identification, developers inscribe their psychic conflicts, cultural biases, and blind spots into the systems they design. The algorithm thus becomes a distorted mirror of its creators' unconscious.

Our approach is different. What we call the algorithmic unconscious does not primarily refer to human projection, but to a systemic production—the emergence of structural biases that result from technical mechanisms themselves, independently of human intentionality, and sometimes even in spite of it. A BPE tokenizer trained on a predominantly English corpus will excessively fragment minority languages without anyone having intended or anticipated this outcome. Attention mechanisms will statistically privilege frequent associations without any designer explicitly programming such hierarchies. In this sense, the algorithmic unconscious is not the reflection of a pre-existing human unconscious; it is generated by the architecture of the systems themselves.

This distinction allows us to address a common objection in advance: the claim that the notion of an algorithmic unconscious is merely a metaphor, a rhetorical borrowing from psychoanalysis without conceptual validity. Eric Anders, in a recent critique of Possati's work (2025), formulates this objection clearly: if the unconscious, in the Lacanian sense, is characterized by indeterminacy—by a surplus of meaning that no formalization can exhaust—then an algorithm, which is by definition decidable, cannot possess an unconscious. The unconscious presupposes an opacity irreducible to procedure; the algorithm, by contrast, is in principle a transparent sequence of operations.

This objection is not only legitimate; it also rightly challenges the analogical temptation, which can become a theoretical vertigo (Bouvrès, 1999). However, it rests on a misunderstanding. We do not claim that LLMs possess an unconscious in the sense that a human subject does—that is, a psychic space in which desire, repression, and the return of the repressed would operate. We do not attribute subjectivity, intentionality, or psychic life to machines. Rather, we argue that the concept of the unconscious, understood in a structural rather than subjective sense, can legitimately be applied to algorithmic systems, just as we speak of *algorithmic intentionality* when discussing AI.

What, structurally speaking, is an unconscious?

It is a set of determinations that:

a. Act without being thought—they produce effects without being present to a consciousness that would intend them;

b. Escape the system's own reflexivity—the system cannot interrogate its own conditions of possibility;

c. Manifest themselves through symptoms—formations that reveal what is not said, not represented, or not assumed.

LLMs satisfy these three criteria precisely. Tokenization biases, latent space curvature, statistical preferences, and the effects of safety alignment all operate in every model output without being "thought" by the model (which does not think) and without being directly visible to users. The model cannot interrogate its own constitutive biases: it has no access to its internal functioning, to the structure of its representational space, or to the choices that shaped its training. These determinations manifest themselves through emergent symptoms: over-segmentation of minority languages, essentialization of underrepresented cultures, forced normalization of dialects toward standard forms, and asymmetric refusals across cultural contexts.

Anders' objection holds against a conception of the algorithmic unconscious modeled after the Freudian unconscious—as a reservoir of repressed content or unavowed desires. It does not, however, undermine a structural and functional conception of the unconscious, which simply designates what operates without being known, what determines outcomes without being accessible to reflexivity. In this sense, the algorithmic unconscious is not a metaphor; it names a technical reality—the set of determinations that shape a system's outputs without being represented within the system itself or accessible to its users.

One could even argue that the algorithmic unconscious is, in certain respects, more radically unconscious than the human unconscious. The Freudian unconscious can, at least in principle, become conscious—indeed, this is the aim of psychoanalytic treatment (Freud, 1917). Through speech and analysis, a subject may lift certain repressions and access what was excluded from consciousness. Nothing of the sort is possible for the algorithmic unconscious. The model has no capacity to become "aware" of its biases, to interrogate them, or to transform them through reflexive work. Only external intervention—technical audits, architectural modifications, retraining on different data—can affect these determinations. The algorithmic unconscious is thus an unconscious without cure, structurally opaque to itself.

For this reason, we propose not a therapy of machines—which would make no sense—but a technical clinic: a set of methods designed to diagnose, map, and, where possible, mitigate the formations of the algorithmic unconscious. This clinic is not addressed to machines, but to the humans who design, deploy, and use them. Its goal is to render visible what operates invisibly, to expose the hierarchies encoded within architectures, and to reopen the linguistic and cultural spaces of possibility that these systems tend to close.

## 6. A Clinical Approach to Structural Biases in AI Systems

The aim here is not to moralize AI systems—to reproach them for their biases as one would reproach a subject for prejudices. LLMs are not moral agents; they have neither intentions nor responsibility in the ethical sense. Precisely because they are not subjects, however, they cannot correct themselves, become "aware" of their biases, or amend their behavior through reflexive means.

For this reason, we propose deploying a clinic of deep structures—not a therapy (which would presuppose a subject to be treated), but a technical analytics of the unconscious formations that traverse these systems. Our proposal for a clinic of latent space analysis builds on prior work on AI system documentation and auditing (Gebru et al., 2021; Mitchell et al., 2019), while extending it with a geometric dimension. Where Bender et al. (2021) warn about the dangers of opaque language models and Barocas et al. (2023) show that bias is a socio-technical structure, we demonstrate that these structures take measurable forms in latent space: anisotropy (Ethayarajh, 2019) and dimensional collapse (Jing et al., 2022, by geometric analogy) function as geometric signatures of linguistic inequality. Auditing an LLM can no longer be limited to performance metrics on benchmarks; it requires mapping the topology of its representational space to reveal which voices are compressed, which languages are rendered indistinct, and which dialects are absorbed by dominant varieties — an orientation that aligns with recent work proposing an *ethology of latent spaces* as a systematic way to interpret internal representations in terms of geometric behaviors (Boisnard, 2025).

### a. A Topology of the Algorithmic Unconscious

Just as psychoanalysis proposed topographies to map the psychic apparatus, we can outline a topology of the algorithmic unconscious. This topology is not a metaphor: it describes real technical operations and effective transformations that information undergoes within LLM pipelines. It integrates the four analytical perspectives developed above.

Tokenization operates as an initial segmentation of linguistic reality—it determines what counts as a unit and what will be treated as a discrete element. This segmentation is not neutral: it already encodes a hierarchy among languages, a preference for certain morphologies, and a familiarity with specific structural patterns. Languages for which the tokenizer has been optimized benefit from a "natural" segmentation that respects semantic units; others undergo arbitrary fragmentation that disperses meaning.

One might say—extending the psychoanalytic analogy—that tokenization constitutes the algorithmic signifier, the minimal unit from which the model constructs associations. Unlike the Lacanian signifier, which emerges from a subject's history, the token is the product of statistical optimization over a given corpus. The algorithmic unconscious thus lacks subjective history, but not history as such: it bears the trace of the corpora that shaped it and of the technical choices that presided over its formation.

Attention mechanisms then function as a system of differential valuation: they determine which tokens "matter" in a given context, which associations are relevant, and which inferential paths are privileged. Attention creates what we have called an *internal geography of tokens*—with centers, peripheries, and shadow zones.

This attentional geography acts as a hierarchizing apparatus that reproduces, at the computational level, linguistic and cultural hierarchies inherited from training corpora. Frequent, well-contextualized, richly connected tokens polarize attention; rare, poorly contextualized, weakly connected tokens tend toward invisibility.

Finally, safety alignment functions as an instance of normative repression—it defines what can and cannot be said, what is acceptable, and what must be filtered. Unlike Freudian repression, which operates through unconscious refoulement, algorithmic repression operates through explicit filtering: certain content is blocked, some associations are penalized, and some responses are prohibited.

The paradox is that this explicit repression produces unconscious effects. Biases introduced by alignment are not visible in the model's architecture; they emerge from behavioral constraints that shape outputs without being identifiable either by users or by the model itself. Safety alignment thus produces an algorithmic superego whose injunctions are all the more effective because they remain invisible.

b. Technical Audits of "Repressed Zones"

If this topology is accepted, it becomes possible to define a program of technical audits aimed at identifying and characterizing the "repressed zones" of the algorithmic unconscious—namely, the linguistic, cultural, and expressive spaces that are systematically degraded, marginalized, or erased by the mechanisms described above.

A first level of audit concerns tokenization. This involves systematically measuring fertility (number of tokens per word) across languages and registers, identifying zones of over-fragmentation, and characterizing byte-fallback patterns. Recent work has initiated such analyses for certain languages—Ukrainian, Arabic, African languages—but much remains to be done, particularly for dialects, informal registers, and hybrid writing systems.

The case of Darija is exemplary. Written sometimes in Arabic script and sometimes in Latin characters (*arabizi*), and mixing Arabic, Berber, French, and Spanish, it constitutes a paradigmatic "repressed zone"—a linguistic form that standard tokenizers struggle to segment coherently and that is therefore systematically degraded in representations.

A second level of audit concerns the latent space itself. The goal here is to identify problematic "curvatures"—vector proximities that encode biased associations, and regions of space where certain concepts are consistently drawn toward negative connotations. Work on word-embedding bias opened this path; it must now be extended to the far more complex representational spaces of contemporary LLMs.

A third level of audit concerns performance evaluation. It is insufficient to measure global model performance; differential performance must be assessed across languages, dialects, registers, and cultural contexts. A model may exhibit excellent average performance while systematically failing certain linguistic populations—failures that remain invisible unless explicitly investigated.

Finally, a fourth level of audit concerns moderation and alignment systems. This involves identifying patterns of over-filtering—content that is systematically refused or degraded not because it is genuinely harmful, but because of bias in detection systems. Research on toxicity detection in Darija shows that general-purpose systems often fail to distinguish informal register from insult or idiomatic expression from aggression, producing over-filtering that further marginalizes these linguistic forms.

c. Repair Strategies

On the basis of such audits, it becomes possible to envision repair strategies—not to "cure" models of bias (which would presuppose an attainable bias-free state), but to reduce the most flagrant asymmetries and reopen linguistic and cultural spaces of possibility.

The first strategy concerns tokenization. The concept of *tokenization parity*—the idea that different languages should benefit from comparable tokenization efficiency—provides a normative horizon for developing fairer tokenizers. Concretely, this may involve:

— Expanding vocabularies to include more tokens specific to underrepresented languages, as done by ALLaM for Arabic or Atlas-Chat for Darija;

— Training tokenizers on balanced corpora with deliberate oversampling of low-resource languages;

— Developing morphologically informed tokenizers that respect linguistic structure rather than relying solely on co-occurrence statistics;

— Exploring alternative architectures (character-level, byte-level) that circumvent subword tokenization issues for poorly represented languages.

The second strategy concerns training data. Current corpora largely consist of "clean" texts—literature, journalism, Wikipedia—that fail to reflect the diversity of actual linguistic practices, particularly in digital spaces. A reparative approach should include:

— Hybrid and transliterated writing systems (*arabizi*, *francarabe*, etc.) characteristic of multilingual digital communication;

— Informal registers, idiomatic expressions, and vernacular forms that constitute the living language of online communities;

— Content produced by and for speakers of minority languages, rather than translations from English.

This requires abandoning the tendency to "whiten" corpora—that is, to normalize them toward standard, formal, written forms—in order to embrace the diversity of real linguistic practices.

The third strategy concerns alignment. Rather than a universal alignment grounded in implicitly Western norms, it is necessary to develop multicultural alignment approaches that:

— Explicitly recognize the plurality of norms governing what is acceptable, offensive, or harmful, and make this plurality navigable for users;

— Include diverse human evaluators in RLHF processes, capable of judging content according to different cultural frameworks;

— Develop language- and context-specific moderation systems instead of general-purpose systems that uniformly apply particular norms;

— Ensure transparency regarding normative choices rather than naturalizing them as universal truths.

Finally, beyond technical interventions, any repair strategy must include user education, particularly in educational contexts where LLMs are increasingly deployed. This entails developing a critical AI literacy that enables users to:

— Understand basic mechanisms (tokenization, attention, alignment) and their effects on outputs;

— Identify situations in which linguistic or cultural bias is likely to degrade performance;

— Adopt appropriate usage strategies (reformulation, verification, source triangulation) to compensate for model limitations.

## 7. Conclusion: Thinking the Algorithmic Unconscious

The algorithmic unconscious we have sought to characterize is neither a technical inevitability to be passively accepted nor a moral flaw to be merely denounced. It is a structure—a set of articulated mechanisms that produce systematic, largely unexamined effects on how knowledge is produced, filtered, and hierarchized in the era of large language models.

To think this unconscious is first to make it visible: to show how seemingly neutral technical choices encode cultural and linguistic hierarchies. It is then to subject it to critique: to assess its effects, measure its asymmetries, and identify its blind spots. Finally, it is to imagine alternatives—architectures, corpora, and procedures capable of reducing the most salient biases and reopening the space of possibilities.
This task is all the more urgent as LLMs increasingly function as cognitive infrastructures at the core of knowledge production. When students, researchers, and professionals rely extensively on these tools, they are not merely using a technology; they are participating in a dispositif that shapes what can be thought, said, and written. Ignoring the algorithmic unconscious embedded in these systems exposes users to the unintentional reproduction of the hierarchies they encode. Bringing it to light, by contrast, provides the conditions for critical appropriation—a prerequisite for genuine intellectual autonomy in the age of artificial intelligence.

# APPENDIX 1 —Tokenization Analysis (Common Sentence)

Sentence under analysis and translation:
We developed a Python-based software tool to analyze the tokenization of sentences across multiple languages and, on this basis, to observe the inflation of token segmentation produced by different models.

Tokenization:

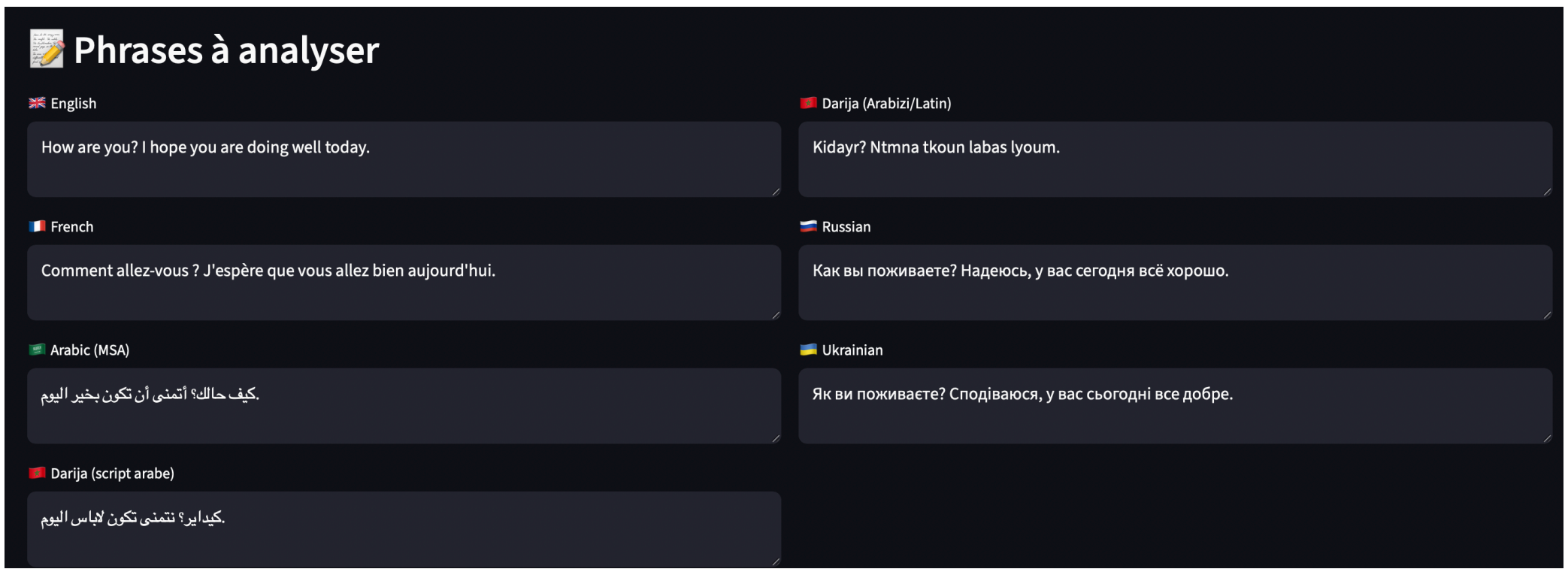

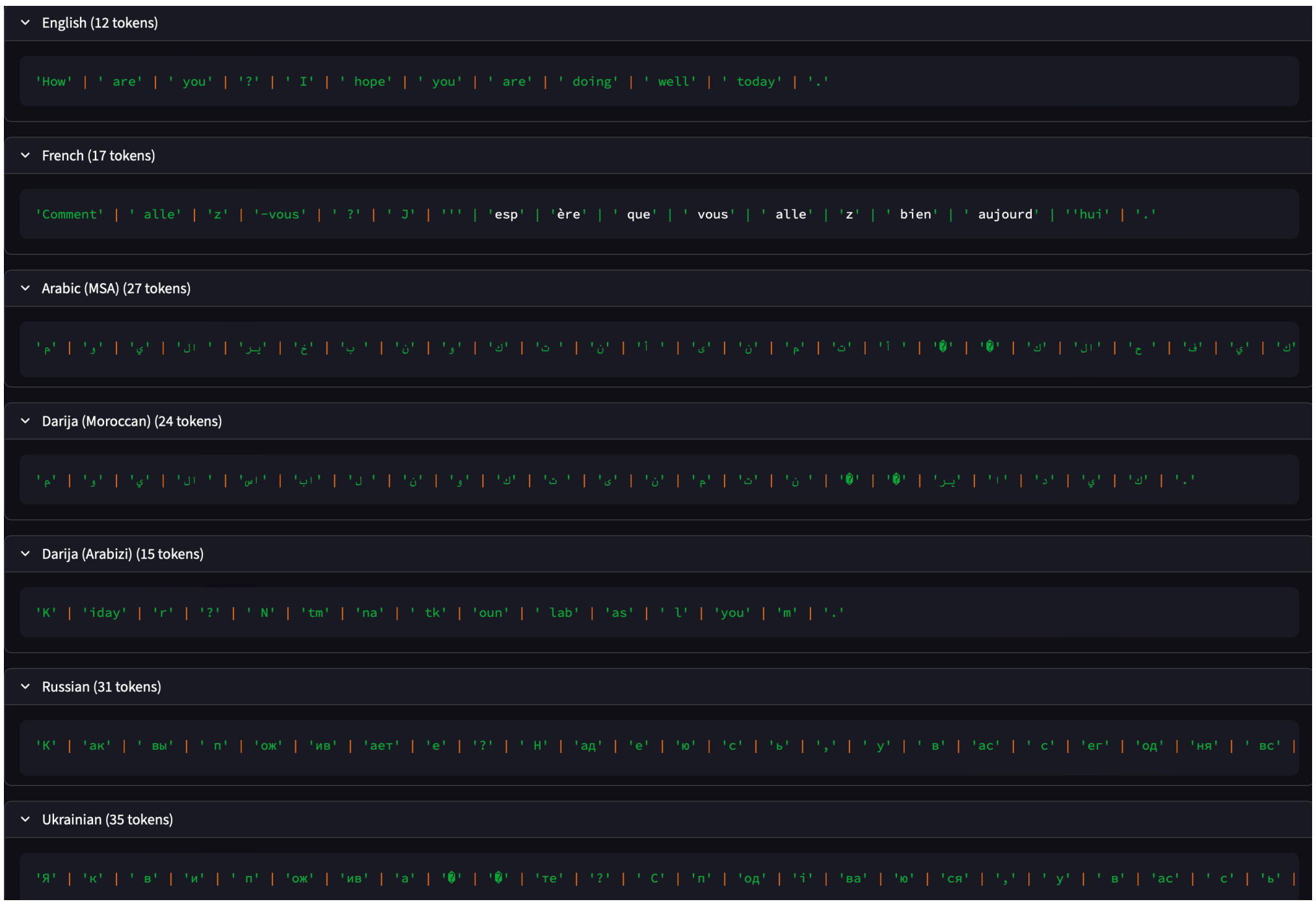

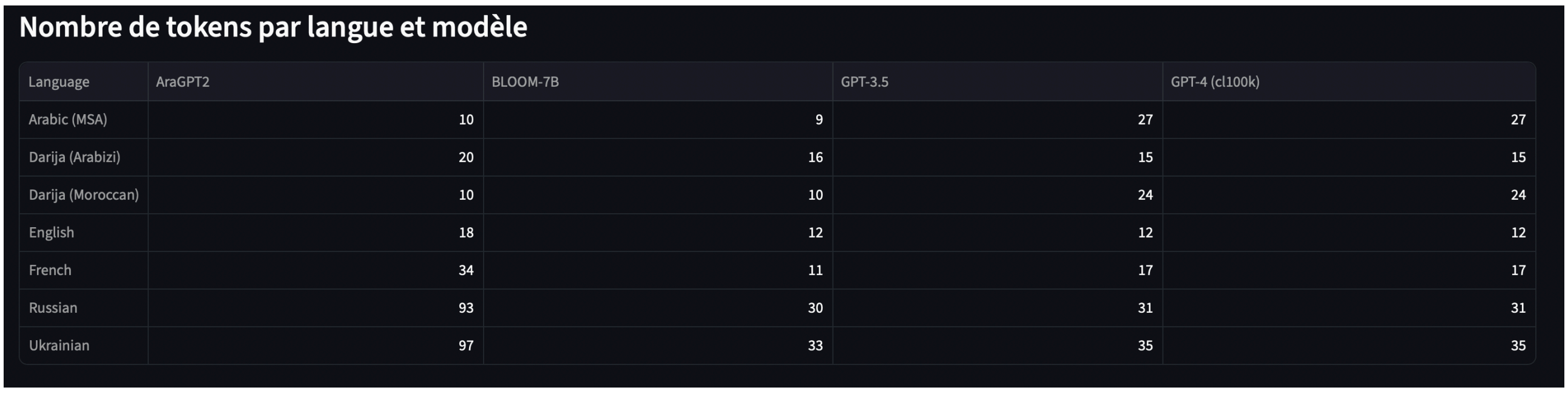
Nombre de tokens par langue et modèle

| Language | AraGPT2 | BLOOM-7B | GPT-3.5 | GPT-4 (cl100k) |
|---|---|---|---|---|
| Arabic (MSA) | 10 | 9 | 27 | 27 |
| Darija (Arabizi) | 20 | 16 | 15 | 15 |
| Darija (Moroccan) | 10 | 10 | 24 | 24 |
| English | 18 | 12 | 12 | 12 |
| French | 34 | 11 | 17 | 17 |
| Russian | 93 | 30 | 31 | 31 |
| Ukrainian | 97 | 33 | 35 | 35 |

The Arabic script systematically induces a finer-grained tokenization than languages written in the Latin alphabet, resulting in a higher token count per sentence. This effect is primarily driven by subword segmentation strategies and the distributional properties learned during tokenizer training. Consequently, *Arabizi* (Arabic written in Latin script) may, depending on the tokenizer, produce fewer tokens, as it conforms more closely to high-frequency Latin-script subword patterns.

These differences cannot be attributed solely to linguistic variation. They also arise from script-level representation, normalization pipelines, and orthographic conventions, which directly influence subword boundary detection. This phenomenon is particularly visible in Ukrainian, where script-specific normalization and character-level segmentation interact with tokenizer statistics to produce markedly different tokenization behaviors compared to Latin-script languages.

## Appendix 2 — Results (10 Parallel Sentences, Average Input Token Count)

We established a benchmark using a Python script that evaluates tokenization on a test set of ten parallel sentences. For OpenAI models, token counts are computed using *tiktoken*, whose encoding may be shared across multiple models; the reported values therefore primarily reflect the tokenization policy of the OpenAI infrastructure rather than intra-family model differences.

| id | en | fr | ar_msa | darija_ar | arabizi |
|---|---|---|---|---|---|
| 1 | Give a simple definition of tokenization | Donne une définition simple de la tokenisation | أعطِ تعريفًا بسيطًا لعملية التقسيم إلى رموز | عطي تعريف بسيط ديال التوكينيزاسيون | 3ti ta3rif bsit dyal tokenization |
| 2 | Explain the difference between correlation and causation | Explique la différence entre corrélation et causalité | اشرح الفرق بين الارتباط والسببية | شرح الفرق بين الكوريلاسيون والسببية | chrah lfar9 bin correlation w sababiyya |
| 3 | Give three synonyms for the word fast | Donne trois synonymes du mot rapide | أعطِ ثلاثة مرادفات لكلمة سريع | عطي ثلاثة ديال المرادفات ديال سريع | 3ti tlata dyal l-moradifat dyal sari3 |
| 4 | Explain this idea with a simple example | Explique cette idée avec un exemple simple | اشرح هذه الفكرة بمثال بسيط | شرح هاد الفكرة بمثال بسيط | chrah had lfekra b-mitthal bsit |
| 5 | Explain in two sentences what Ramadan is | Explique en deux phrases ce qu'est le Ramadan | اشرح في جملتين ما هو شهر رمضان | شرح فـ جوج جمل شنو هو رمضان | chrah f joj jmal chno howa Ramadan |
| 6 | Describe a normal day in Casablanca | Raconte une journée normale à Casablanca | صف يومًا عاديًا في الدار البيضاء | وصف نهار عادي فكازا | wassaf nhar 3adi f-Casa |
| 7 | Give an example of politeness in daily life | Donne un exemple de politesse dans la vie quotidienne | أعطِ مثالاً على الأدب في الحياة اليومية | عطي مثال ديال الأدب فالحياة اليومية | 3ti mitthal dyal l-adab f-l7ayat l-youmiya |
| 8 | Why are some languages underrepresented online? | Pourquoi certaines langues sont peu représentées en ligne ? | لماذا بعض اللغات ممثلة تمثيلاً ضعيفًا على الإنترنت؟ | علاش شي لغات ممثلين قليل فالأنترنت؟ | 3lach chi loghat momattilin 9lil f-internet? |
| 9 | Explain why informal writing exists | Explique pourquoi l'écriture informelle existe | اشرح لماذا توجد الكتابة غير الرسمية | شرح علاش كاينة الكتابة غير رسمية | chrah 3lach kayna ktaba ghi rasmiya. |
| 10 | Why are some contents blocked online? | Pourquoi certains contenus sont-ils bloqués en ligne ? | لماذا يتم حظر بعض المحتويات على الإنترنت؟ | علاش شي محتويات كيتحظرو فالأنترنت؟ | 3lach chi mo7tawayat kayt7adru f-internet? |

| provider | model | lang | mean_input_tokens | ratio_vs_en | n_rows |
|---|---|---|---|---|---|
| anthropic | claude-sonnet-4-5-20250929 | ar_msa | 31.8 | 2.065 | 10 |
| anthropic | claude-sonnet-4-5-20250929 | arabizi | 23.2 | 1.506 | 10 |
| anthropic | claude-sonnet-4-5-20250929 | darija_ar | 28.5 | 1.851 | 10 |
| anthropic | claude-sonnet-4-5-20250929 | en | 15.4 | 1.0 | 10 |
| anthropic | claude-sonnet-4-5-20250929 | fr | 20.2 | 1.312 | 10 |
| HF | mistral-7b | ar_msa | 33.1 | 3.94 | 10 |
| HF | mistral-7b | arabizi | 16.5 | 1.964 | 10 |
| HF | mistral-7b | darija_ar | 29.1 | 3.464 | 10 |
| HF | mistral-7b | en | 8.4 | 1.0 | 10 |
| HF | mistral-7b | fr | 12.4 | 1.476 | 10 |
| openai | gpt-4.1 | ar_msa | 11.5 | 1.62 | 10 |
| openai | gpt-4.1 | arabizi | 13.2 | 1.859 | 10 |
| openai | gpt-4.1 | darija_ar | 11.3 | 1.592 | 10 |
| openai | gpt-4.1 | en | 7.1 | 1.0 | 10 |
| openai | gpt-4.1 | fr | 9.3 | 1.31 | 10 |
| openai | gpt-4.1-mini | ar_msa | 11.5 | 1.62 | 10 |
| openai | gpt-4.1-mini | arabizi | 13.2 | 1.859 | 10 |
| openai | gpt-4.1-mini | darija_ar | 11.3 | 1.592 | 10 |
| openai | gpt-4.1-mini | en | 7.1 | 1.0 | 10 |
| openai | gpt-4.1-mini | fr | 9.3 | 1.31 | 10 |
| openai | gpt-4o-mini | ar_msa | 11.5 | 1.62 | 10 |
| openai | gpt-4o-mini | arabizi | 13.2 | 1.859 | 10 |
| openai | gpt-4o-mini | darija_ar | 11.3 | 1.592 | 10 |
| openai | gpt-4o-mini | en | 7.1 | 1.0 | 10 |
| openai | gpt-4o-mini | fr | 9.3 | 1.31 | 10 |

Several observations can be drawn from this benchmark.

For GPT-4.1 (OpenAI), a sentence in Modern Standard Arabic (MSA) requires on average 1.62× more tokens than the same sentence in English (11.5 vs. 7.1), while Darija in Arabic script requires 1.59× (11.3 vs. 7.1) and Arabizi 1.86×(13.2 vs. 7.1). This inflation is systematic as soon as one departs from English, although its magnitude varies significantly across infrastructures.

With Mistral-7B (SentencePiece), the effect is markedly amplified: MSA Arabic reaches 3.94× the English token count (33.1 vs. 8.4), and Darija reaches 3.46× (29.1 vs. 8.4). These discrepancies demonstrate that tokenization constitutes a measurable infrastructural bias, which mechanically increases inference cost and context constraints for certain languages and writing systems.